# Radiology-Report Semantic Modelling and Host-Response Laboratory Biomarkers for Multimodal Survival Prediction in Lung Cancer

Jingxiang Shi[1], Yiming Wang[1], Zhengda Li[3], Yan Zhang[2], Weihua Meng[1], Yuqi Ma[1], Xiaoyan Li[3], Feng-Ming (Spring) Kong[2], Gen Yang[1*]

1 State Key Laboratory of Nuclear Physics and Nuclear Technology, Peking University, Beijing, China;

2 The University of Hong Kong-Shenzhen Hospital, Shenzhen, China;

3 Cancer Hospital of China Medical University, Shenyang, China.

*Correspondence: Gen Yang, gen.yang@pku.edu.cn

## Abstract

Introduction: TNM staging remains central to lung cancer management, but survival can vary substantially among patients assigned to the same anatomic stage. Radiology reports are a routinely available form of biomedical imaging-derived text, encoding radiologist interpretation of tumor extent, lymph-node involvement, metastatic pattern, and longitudinal disease context. These report semantics may complement laboratory measurements that reflect host response, systemic inflammation, nutritional reserve, and organ function.

Methods: We developed a multimodal adaptive risk score (AMRS) in a retrospective two-center cohort of 574 patients with lung cancer diagnosed between 2017 and 2026. Radiology reports were encoded with a domain-adapted MC-BERT branch, clinical and laboratory variables were modeled after Mahalanobis-distance-based imputation with random survival forests, and both branches were combined through weighted risk fusion. Performance was evaluated in training (n = 459) and test (n = 115) cohorts using the C-index, time-dependent AUC, Kaplan-Meier stratification, SHAP analysis, and Cox regression of fused model components.

Results: AMRS achieved C-index values of 0.920 in the training cohort and 0.849 in the test cohort, with time-dependent AUC values approaching 0.97 and 0.88, respectively. The score separated survival trajectories across clinical subgroups and within TNM-related strata. SHAP analysis identified hemoglobin, sodium, hematocrit, lactate dehydrogenase, CA125, CA199, inflammatory-cell measures, coagulation markers, nutritional markers, and age as major contributors. Univariate Cox analysis identified six fused components associated with overall survival.

Conclusion: AMRS provides an interpretable multimodal framework for individualized survival prediction in lung cancer by integrating imaging-derived report semantics with host-response laboratory biomarkers. The model may complement TNM-based risk stratification in imaging-centered oncology workflows. Prospective validation, calibration analysis, ablation testing, and formal clinical-utility assessment are needed before clinical deployment.

## Introduction

Lung cancer remains one of the leading causes of cancer incidence and cancer-related mortality worldwide despite advances in screening, surgery, radiotherapy, targeted therapy, immunotherapy, and supportive care [1,2]. Survival differs not only by histologic subtype and treatment exposure but also by systemic host condition, occult disease burden, functional status, and institutional practice. Prognostic assessment therefore affects treatment selection, follow-up intensity, clinical-trial stratification, and communication with patients.

The TNM system provides the standard anatomic language for lung cancer prognosis and treatment planning [3,4]. Its strength is standardization: tumor size or invasion, nodal involvement, and metastatic spread are summarized in a form that clinicians can interpret and compare. Its limitation is that it compresses biologically and clinically heterogeneous patients into discrete anatomic categories. Patients with the same TNM stage may differ in inflammatory state, nutritional reserve, anemia, organ function, tumor-marker burden, radiographic disease pattern, and treatment tolerance. These non-anatomic dimensions can influence survival but are not fully represented by the stage label [5-7].

Routine clinical workflows already generate data that may help resolve this gap. Radiology reports are not pixel-level images, but they are imaging-derived clinical data: they contain expert interpretation of tumor extent, lymph-node involvement, metastatic patterns, pleural or distant spread, and longitudinal imaging context. This makes report semantics relevant to biomedical imaging analysis and medical-physics-adjacent oncology workflows, especially when original image data are not available. Laboratory tests reflect host physiology, cancer-related inflammation, coagulation, organ function, and nutritional status [7,20-23]. The challenge is to transform heterogeneous imaging-derived text and tabular clinical data into a reliable, interpretable prognostic score.

Artificial intelligence and survival modeling offer a route to integrate these signals [8,9]. However, prior prognostic models often rely on a single modality, single-center data, limited validation, or opaque representations with weak clinical interpretation. For lung cancer prognosis, a clinically useful model should not merely improve a discrimination metric. It should show where the additional information comes from, whether the improvement persists in held-out patients, and whether the model refines categories already used by physicians [10,11,28-35].

This study presents a multimodal adaptive risk score (AMRS) that combines radiology-report semantics with clinical-laboratory information in a two-center lung cancer cohort. The work is positioned as semantic modelling of biomedical imaging reports rather than as image radiomics. The evidentiary sequence begins with construction of the multimodal pipeline, proceeds to benchmark comparison and clinical interpretability, and then tests whether the score remains informative across clinical subgroups and within TNM-related risk strata. The final analysis examines whether the fused latent representation contains components associated with overall survival.

## Methods

### Study design and cohort construction

This retrospective two-center study used clinical data from The University of Hong Kong-Shenzhen Hospital and the Cancer Hospital of China Medical University. The source dataset included electronic medical records, diagnostic and follow-up information, radiology reports, treatment variables, and laboratory measurements for patients diagnosed with lung cancer between 2017 and 2026. Overall survival was the primary endpoint. Patients with missing multimodal inputs or follow-up shorter than one month were excluded.

Among 1129 screened patients, 555 were excluded because of follow-up shorter than one month (n = 283) or missing imaging reports (n = 272). The final cohort included 574 patients with 208 events and was divided into a training set (n = 459; events = 166; median follow-up, 27.2 months) and a test set (n = 115; events = 42; median follow-up, 30.2 months). Baseline variables included age, sex, histology, ECOG performance status, TNM stage, treatment modality, radiology report text, and laboratory biomarkers. The study design reflects a clinically pragmatic setting in which both structured measurements and narrative imaging interpretation are available before survival prediction.

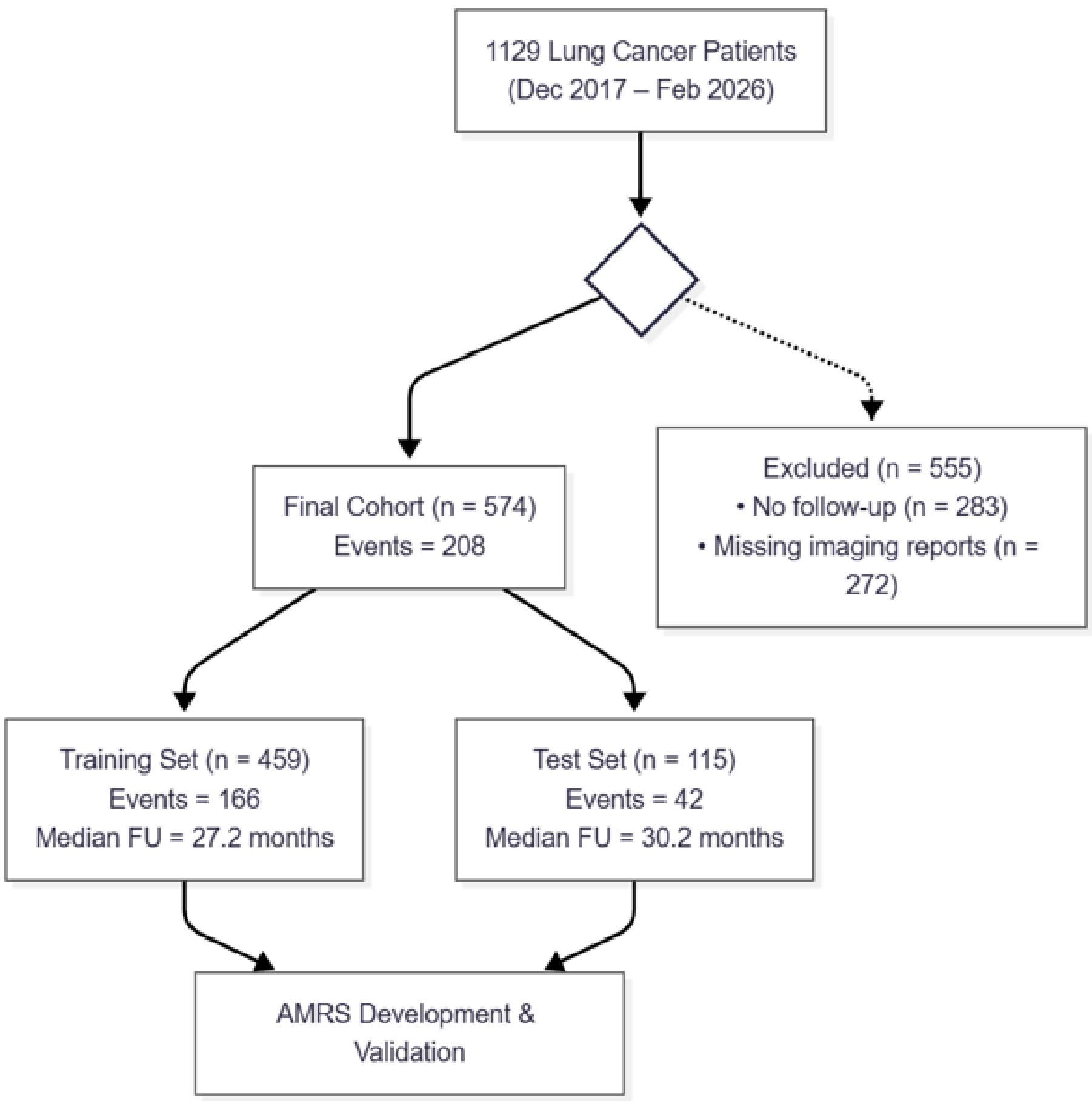

*Figure 1. Cohort inclusion and partitioning. The source cohort included 1129 patients with lung cancer diagnosed between December 2017 and February 2026. After exclusion of patients with short follow-up or missing imaging reports, 574 patients were included for AMRS development and validation.*

## Multimodal AMRS framework

The AMRS model contains two parallel branches that converge into a weighted risk-fusion layer. The imaging-text branch transforms radiology reports into semantic representations designed to capture report-level evidence of tumor distribution, lymph-node enlargement, and imaging-detected metastatic disease. The clinical-laboratory branch transforms structured variables and laboratory measurements into a nonlinear survival-risk estimate. The final score is generated by weighted fusion, allowing the model to combine radiographic disease semantics with physiologic and biochemical risk information.

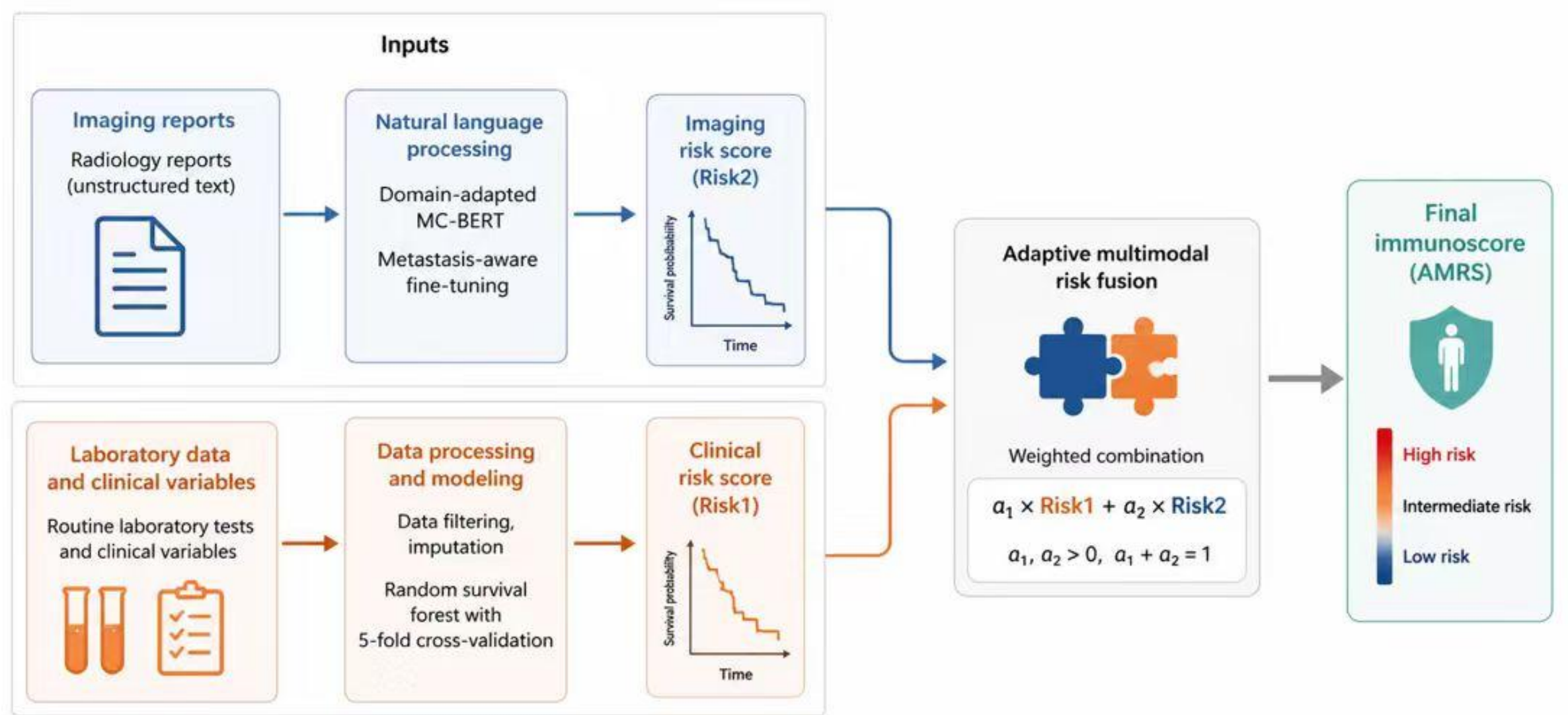

*Figure 2. Multimodal AMRS workflow. Radiology reports are encoded with a domain-adapted MC-BERT branch to produce an imaging-text risk score (Risk2). Clinical variables and laboratory measurements are processed with missing-value imputation and random survival forest modeling to produce a clinical risk score (Risk1). Weighted fusion generates the final patient-level risk score.*

## Radiology report semantic branch

Radiology reports were encoded using a domain-adapted MC-BERT language-modeling pipeline. The objective was not to replace radiologist interpretation, but to convert clinically meaningful report semantics into a quantitative survival-risk representation. This design was informed by transformer language models, biomedical or clinical BERT variants, and recent work on extracting staging information from unstructured lung cancer reports [16-18,32,33]. Report embeddings were reduced using principal component analysis and then filtered using Cox-guided feature selection to derive an imaging-text risk score (Risk2). This branch was intended to capture disease-pattern information that may not be fully represented by structured TNM labels [10,11].

## Clinical-laboratory branch

Clinical and laboratory variables were harmonized across institutions before modeling. Missing laboratory values were imputed using a Mahalanobis-distance-based strategy to preserve multivariable covariance structure rather than treating each measurement independently. A random survival forest model was then trained to estimate a clinical-laboratory risk score (Risk1) [12]. This branch was designed to capture nonlinear prognostic relationships among anemia, electrolyte balance, inflammation, coagulation, tumor-marker burden, organ function, nutritional reserve, age, and treatment-related clinical variables.

## Fusion and statistical analysis

The final AMRS was calculated as AMRS = 0.8226 x Risk1 + 0.1774 x Risk2. Model discrimination was compared with Cox proportional hazards regression, random survival forests, and FSSVM using the C-index and time-dependent AUC [12,13,31]. Kaplan-Meier curves were used to visualize survival separation by AMRS-defined risk groups across the full cohort, clinical subgroups, and TNM-related groups. Model interpretability was assessed using SHAP-based feature importance [19]. Univariate Cox regression was applied to 32 fused AMRS components and visualized as P-value ranking and hazard-ratio forest plots.

## Results

### Cohort characteristics

The cohort included 574 patients with a median age of 64 years (IQR, 56-71 years); 58% were male. Histologic distribution included lung adenocarcinoma (46%), lung squamous cell carcinoma (34%), and small cell lung cancer (20%). ECOG performance status was 0-1 in 72% and 2-3 in 28%. Surgery was performed in 45% of patients, and median follow-up was 30.2 months. The dataset therefore represents a clinically heterogeneous lung cancer population spanning histology, functional status, treatment modality, and institution.

This heterogeneity provided the basis for testing whether a multimodal score could recover prognostic information dispersed across stage, functional status, treatment exposure, imaging interpretation, and laboratory physiology. AMRS was evaluated in a stepwise manner: survival discrimination against conventional benchmarks, clinical signals contributing to the score, and risk separation within clinically familiar subgroups and TNM-related strata.

### AMRS improved survival discrimination compared with benchmark models

AMRS was compared with established survival-modeling benchmarks to determine whether multimodal fusion translated into measurable predictive gain. In the training cohort, AMRS achieved a C-index of 0.920, compared with 0.915 for random survival forests, 0.730 for FSSVM, and 0.716 for Cox proportional hazards regression. The same ordering was largely preserved in the test cohort, where AMRS retained the highest C-index (0.849), followed by random survival forests (0.846), Cox proportional hazards regression (0.820), and FSSVM (0.798). These results indicate that AMRS performed at least as well as the strongest benchmark and exceeded the conventional proportional hazards model. The narrow margin over random survival forests suggests that the clinical-laboratory branch contributed substantially to the final score.

The temporal performance profile further supported this pattern. AMRS maintained high time-dependent AUC across serial survival horizons, with training values approaching 0.97 and test values remaining around 0.88. This stability is important because clinical use requires risk separation over time, not only at a single prespecified endpoint. At the same time, the closeness between AMRS and random survival forests in the held-out cohort defines a key revision priority: confidence intervals, paired model comparisons, and ablation analyses are needed to quantify the independent value added by the imaging-text branch.

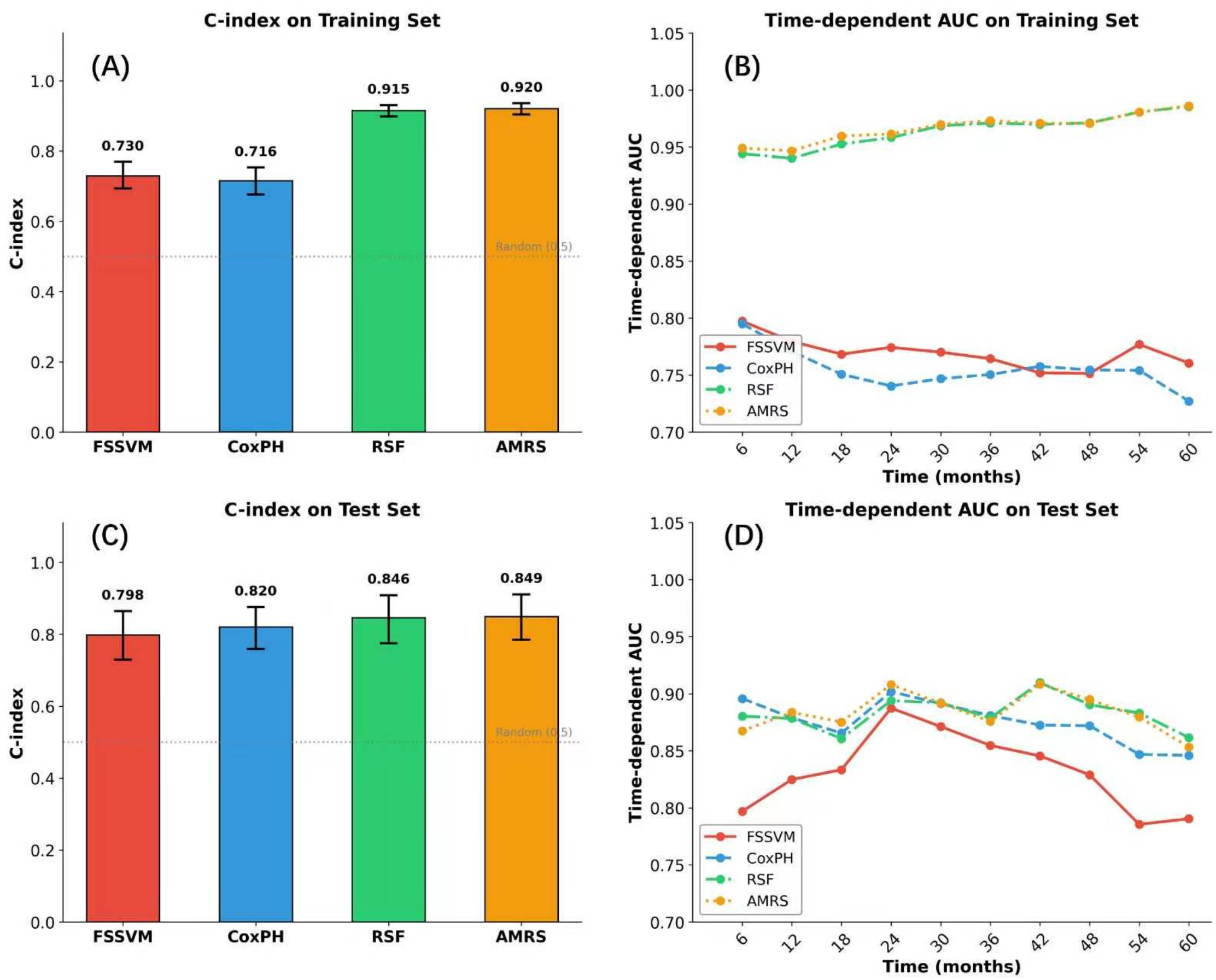

*Figure 3. Discrimination of AMRS and benchmark survival models. C-index values are reported for the training and test cohorts. Time-dependent AUC curves summarize discrimination across follow-up intervals for FSSVM, Cox proportional hazards regression, random survival forests, and AMRS.*

## Clinical-laboratory feature importance connected AMRS risk to plausible biology

After discrimination had been established, the clinical-laboratory branch was examined for interpretability and clinical coherence. SHAP-based ranking identified hemoglobin concentration as the most influential feature, followed by sodium, hematocrit, lactate dehydrogenase, CA125, CA199, monocyte absolute count, PDW-CV, total protein, fibrinogen, age, albumin, urea, platelet count, ALT, lymphocyte absolute count, CEA, MCHC, creatinine, MCV, magnesium, RBC, WBC, sex, D-dimer, MCH, PT, and GGT. The distribution of these variables suggests that the model drew prognostic information from anemia, electrolyte disturbance, tissue injury, tumor-marker burden, systemic inflammation, coagulation activation, nutritional reserve, renal and hepatic function, and general physiologic vulnerability [20-27].

This pattern helps bridge model output and clinical reasoning. A high-risk AMRS profile was not driven by a single isolated biomarker, but by a constellation of abnormalities consistent with impaired physiologic reserve and more aggressive systemic presentation. Conversely, lower-risk profiles were more consistent with preserved blood indices, nutritional markers, and organ-function measures. Thus, feature-importance analysis supports the biological plausibility of the clinical-laboratory branch while remaining appropriately descriptive. SHAP importance identifies influential predictors, but it does not establish causal mechanisms [19].

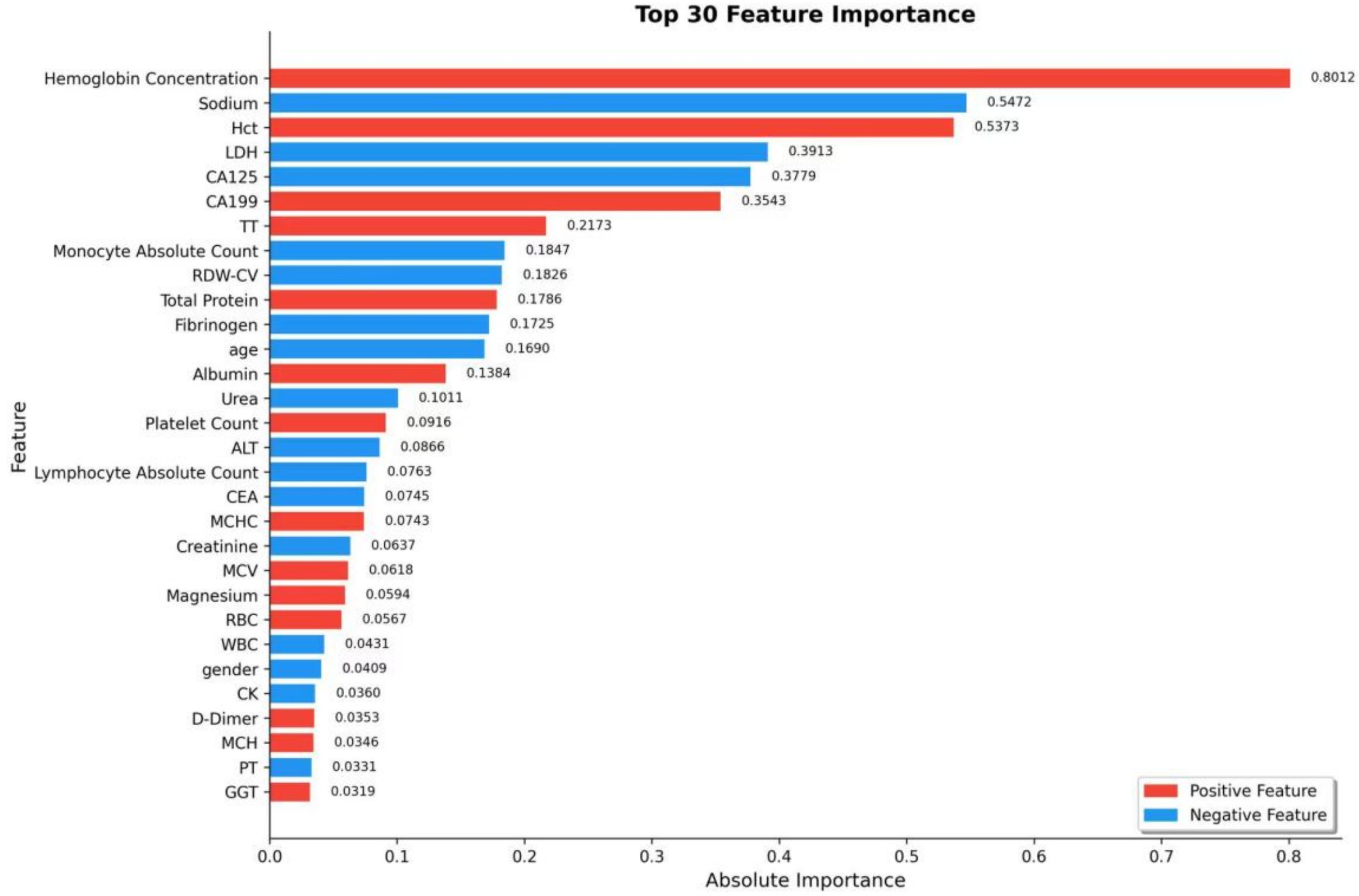

*Figure 4. Clinical-laboratory feature importance. SHAP-based ranking of variables in the clinical risk branch, including hematologic indices, electrolytes, tumor markers, inflammatory variables, coagulation markers, nutritional markers, organ-function measures, and age.*

## AMRS separated survival across clinical subgroups

The next analysis tested whether AMRS remained informative beyond the overall cohort. A clinically useful prognostic score should not depend entirely on one dominant subgroup or one institutional distribution. It should preserve risk separation when patients are viewed through categories that clinicians already use. Kaplan-Meier analyses showed separation between AMRS-defined high- and low-risk groups across subgroup panels, including strata defined by stage, histology, ECOG performance status, treatment modality, and institutional source.

The implication is that AMRS may act as an additional risk layer rather than a substitute for established clinical descriptors. Once a patient has been categorized by histology, performance status, or treatment context, the score may still identify patients whose survival trajectory diverges from others in the same clinical category. This subgroup behavior is encouraging, but the evidence should be made more transparent by adding risk tables, subgroup sample sizes, log-rank P values, and hazard ratios for each panel.

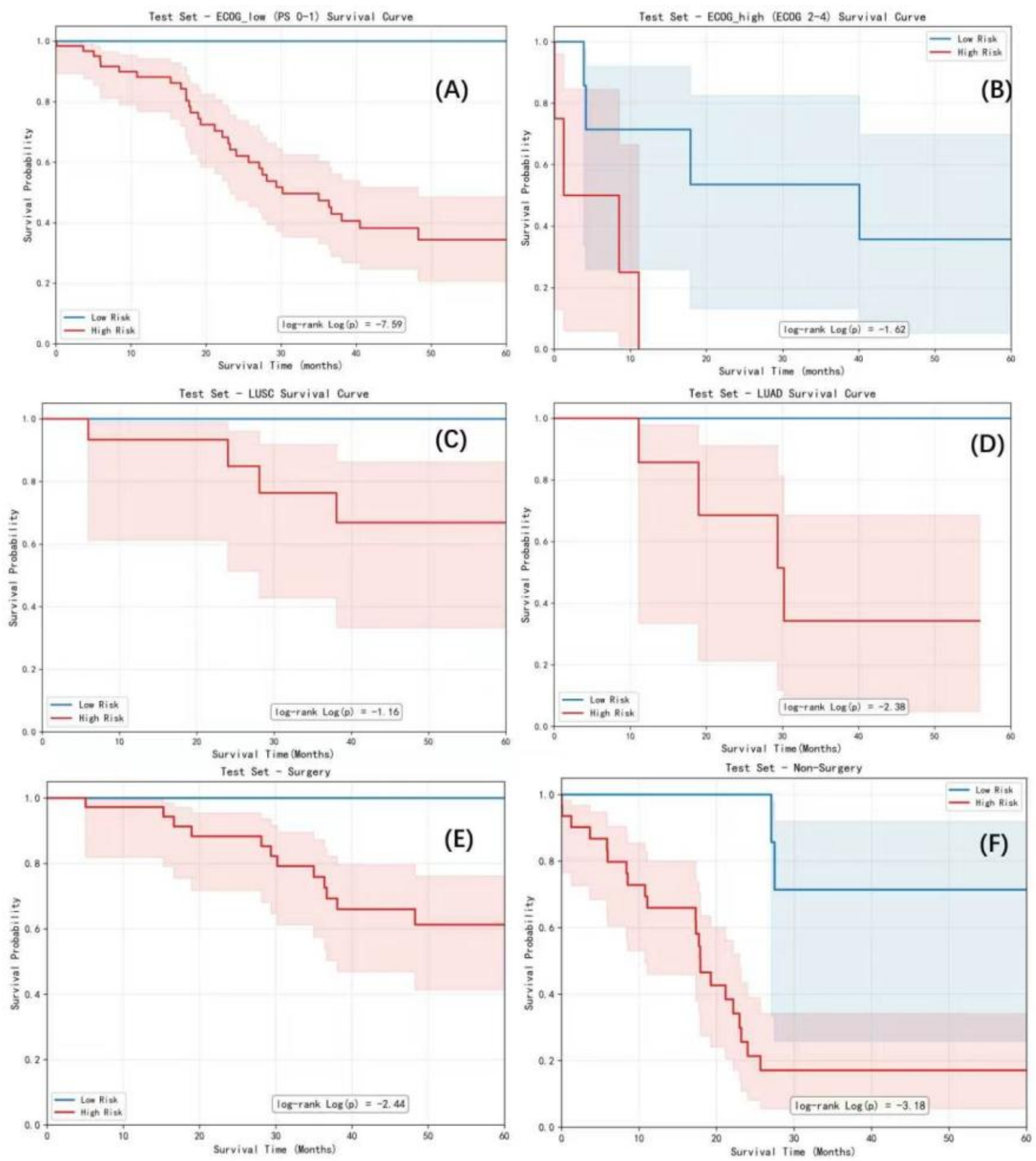

*Figure 5. Survival stratification across clinical subgroups. Kaplan-Meier curves for AMRS-defined high- and low-risk groups across multiple clinical contexts.*

## AMRS refined TNM-related risk stratification

Because the central motivation for AMRS was the residual heterogeneity left by TNM staging, the analysis then focused on TNM-related risk refinement. Patients in Group 1, defined by absence of lymph-node enlargement and absence of imaging-detected metastasis, showed a more favorable survival pattern. Group 3, characterized by imaging-detected metastasis, showed poorer outcomes despite overlap in conventional staging context. Within these TNM-related contexts, AMRS-based stratification separated survival trajectories, suggesting that the multimodal score captures prognostic variation not fully summarized by anatomic descriptors alone.

This is the most clinically relevant part of the Results because it connects the model to a concrete staging problem: two patients can share an anatomic stage label but differ in radiographic disease pattern and systemic physiologic reserve. AMRS appears to recover part of this hidden variation. However, this claim should remain restrained until the manuscript reports the number of patients in each group, the TNM-stage composition of Group 1 and Group 3, the cutoff used to define AMRS risk groups, log-rank P values, and stage-adjusted Cox models.

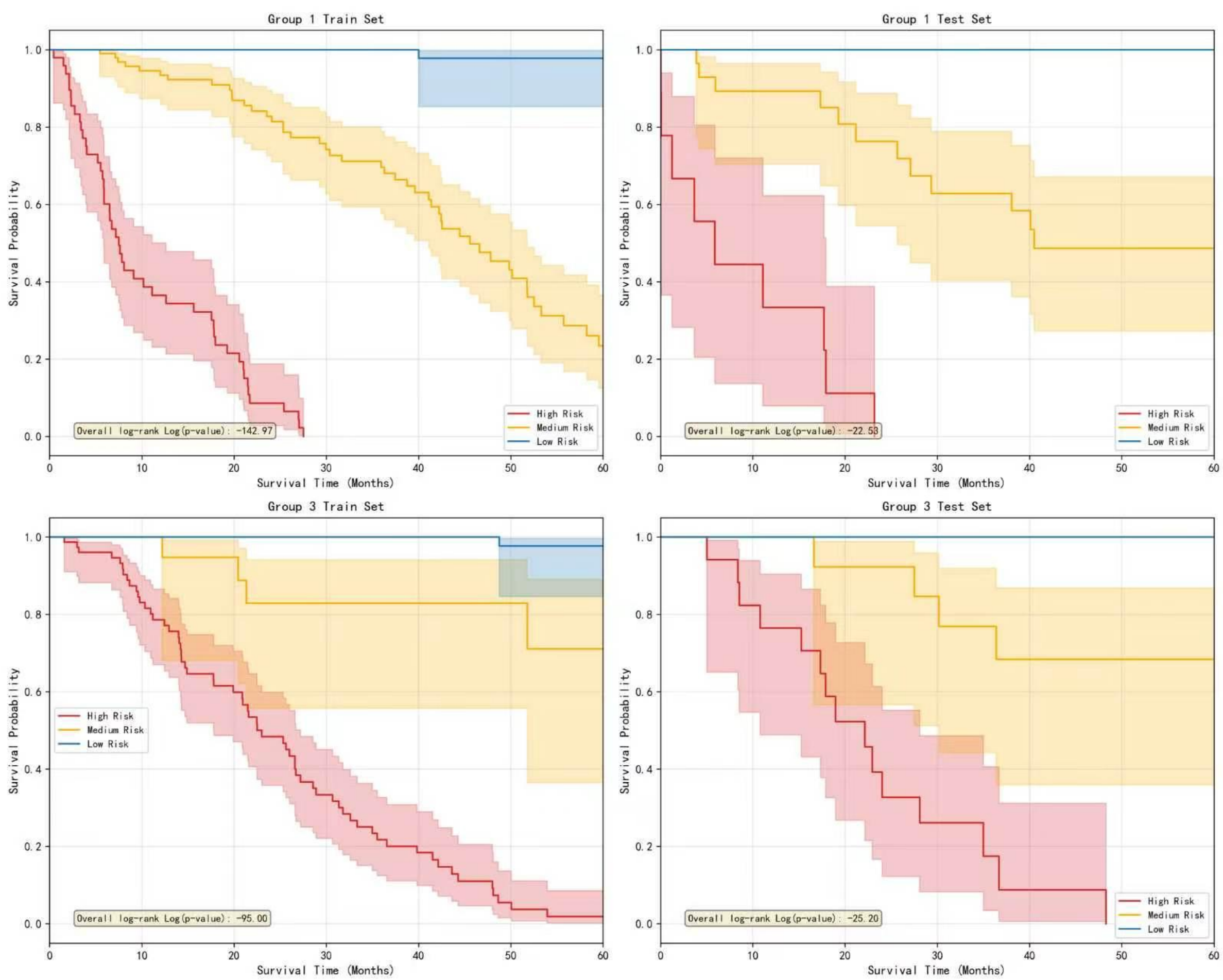

*Figure 6. TNM-related risk refinement. Kaplan-Meier curves for AMRS-defined risk groups within Group 1 and Group 3.*

## Fused AMRS components were associated with overall survival

The final analysis examined whether the learned fused representation contained components individually associated with survival. Among 32 fused components, six reached P < 0.05 in univariate Cox regression: fused_0 (P = 0.0105), fused_1 (P = 0.0179), fused_2 (P = 0.0264), fused_3 (P = 0.0284), fused_4 (P = 0.0410), and fused_5 (P = 0.0437). The corresponding hazard-ratio estimates showed both protective and adverse directions, including fused_0 (HR, 0.762; 95% CI, 0.619-0.938), fused_1 (HR, 1.047; 95% CI, 1.008-1.087), fused_2 (HR, 1.334; 95% CI, 1.034-1.720), fused_3 (HR, 1.316; 95% CI, 1.029-1.683), fused_4 (HR, 0.970; 95% CI, 0.941-0.999), and fused_5 (HR, 1.019; 95% CI, 1.001-1.038).

These associations suggest that the fusion layer encodes more than a single undifferentiated risk signal. Multiple latent dimensions appear to contribute to survival variation, some in protective and others in adverse directions. This finding adds statistical support for the fused representation, but it should not be overinterpreted. Fused_0 to fused_5 are model-derived latent features rather than direct biological biomarkers. Demonstrating independent prognostic value will require multivariable Cox models adjusted for TNM stage, age, ECOG performance status, histology, treatment modality, and institution.

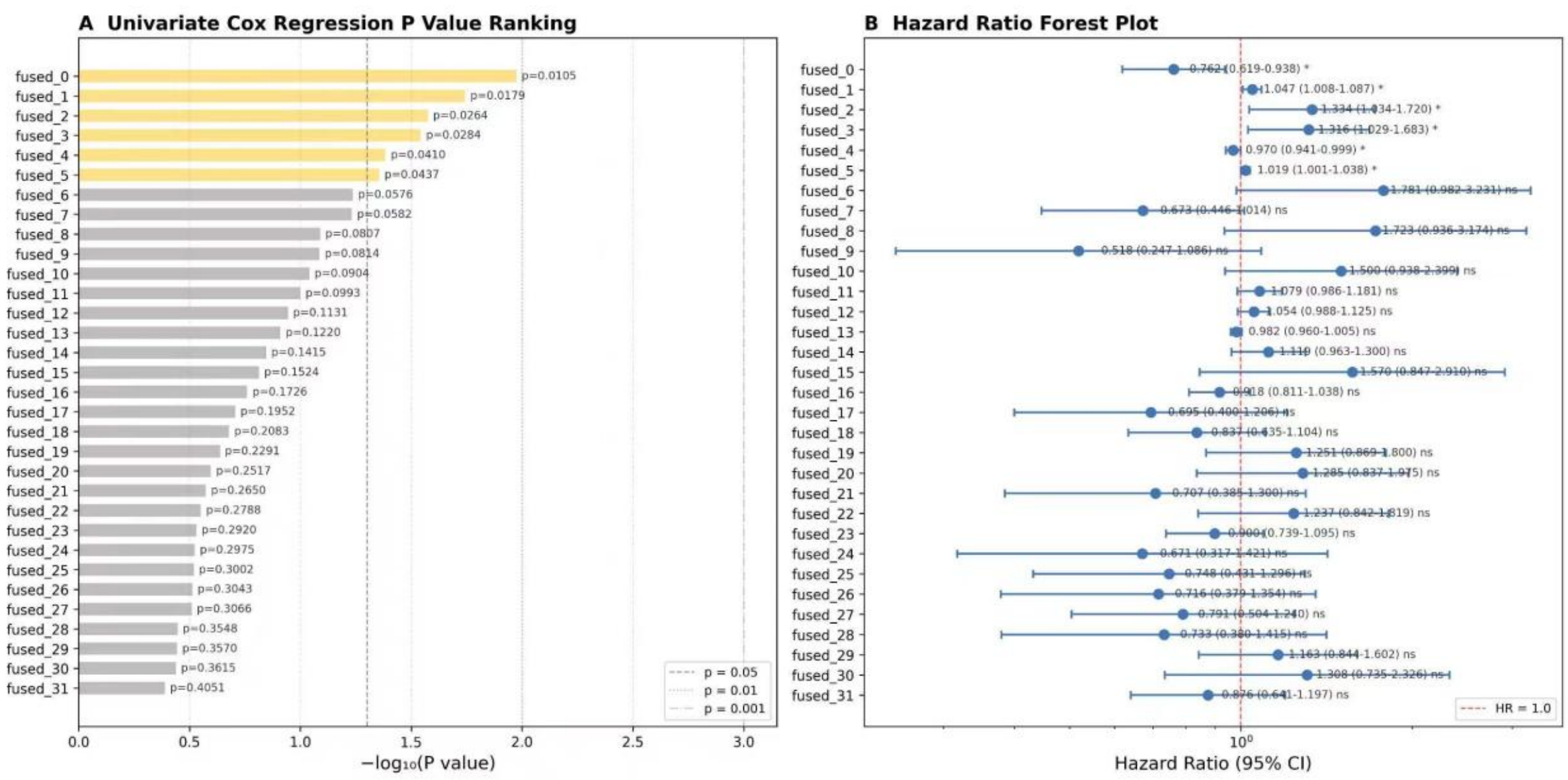

*Figure 7. Cox analysis of fused AMRS components. Univariate Cox regression P-value ranking and hazard-ratio forest plot for 32 fused components.*

# Discussion

This study presents AMRS as a clinically oriented multimodal survival model rather than a stand-alone algorithmic exercise. The main finding is that radiology report semantics and clinical-laboratory variables can be combined to produce a patient-level risk score with strong discrimination in a two-center retrospective cohort. The evidence chain moves from model construction to benchmark performance, clinical interpretability, subgroup survival separation, TNM-related refinement, and fused-feature survival association.

The strongest clinical rationale for AMRS is that it uses data already generated in routine care. Radiology reports encode disease distribution and radiologist judgment, while laboratory measurements reflect physiologic reserve, inflammation, coagulation, tumor-marker burden, and organ function. The integration of these sources is particularly relevant in lung cancer, where anatomic staging is necessary but insufficient to explain individual survival heterogeneity [3-7]. AMRS therefore functions as an additive risk layer that may complement, rather than replace, TNM staging.

From a medical physics perspective, the imaging component of AMRS should be interpreted as radiology-report semantic modelling rather than direct image-feature extraction. This distinction is important because the model does not process voxel-level CT or PET data, dose distributions, or radiomics features. Instead, it uses the radiologist's structured and narrative interpretation as an imaging-derived signal. This design may be useful in imaging-centered oncology workflows where reports are available across institutions but original image series, scanner harmonization, and radiomics preprocessing pipelines are not consistently accessible.

The performance results are encouraging, but they also define the main revision priorities. AMRS outperformed benchmark models numerically, yet the test-set C-index was very close to that of random survival forests. The manuscript should therefore avoid overstating the benefit of multimodal fusion until ablation analysis confirms the incremental value of the imaging-text branch. A stronger final version should include clinical-only, text-only, and fused AMRS comparisons; confidence intervals; paired discrimination tests; calibration curves; and decision-curve analysis. These additions would also align the work more closely with current reporting expectations for machine-learning prediction models [10,28-31,35].

The interpretability findings are clinically plausible. Hemoglobin, hematocrit, albumin, total protein, inflammatory cell counts, fibrinogen, D-dimer, lactate dehydrogenase, tumor markers, renal and hepatic function markers, and age are consistent with known prognostic dimensions in cancer [20-27]. The fused-feature Cox analysis further suggests that survival information is distributed across multiple learned components. Nevertheless, SHAP values and latent-component Cox associations should be interpreted as explanatory aids, not causal evidence.

The TNM refinement analysis is the most clinically important section and the part requiring the most statistical strengthening. The current curves suggest that AMRS can separate patients with different outcomes within TNM-related contexts. To establish this claim convincingly, future versions should report risk cutoffs, group sizes, log-rank P values, adjusted hazard ratios, and whether AMRS remains prognostic after controlling for TNM stage, ECOG performance status, histology, treatment modality, age, and institution.

This study has limitations. It is retrospective and includes two institutions but does not yet provide a fully independent prospective validation cohort. Follow-up maturity, event count, censoring distribution, missingness rate, and model calibration require complete reporting. The model does not incorporate genomic alterations, molecular treatment targets, detailed systemic therapy regimens, original imaging radiomics, or longitudinal post-treatment laboratory dynamics. These limitations do not negate the value of AMRS, but they define the boundary of the current claim and the analyses needed before clinical deployment.

## Conclusion

AMRS integrates radiology-report semantics and clinical-laboratory data to improve individualized survival prediction in lung cancer. By connecting model performance, feature interpretation, subgroup survival separation, TNM-related refinement, and fused-feature Cox associations, the manuscript presents a clearer evidence chain for multimodal risk modeling in imaging-centered oncology workflows. The findings support AMRS as a promising complement to TNM staging, while prospective validation, ablation analysis, calibration, and clinical-utility testing remain essential next steps.